\def\papertitle{Streamable neural audio synthesis with non-causal convolutions}
\def\paperauthorA{Antoine Caillon}
\def\paperauthorB{Philippe Esling}
\def\paperauthorC{Author Three}
\def\paperauthorD{Author Four}

\documentclass[twoside,a4paper]{article}
\usepackage{etoolbox}
\usepackage{dafx_22a}
\usepackage{amsmath,amssymb,amsfonts,amsthm}
\usepackage{euscript}
\usepackage[T1]{fontenc}
\usepackage[utf8]{inputenc}
\usepackage{nimbusserif}
\usepackage{ifpdf}
\usepackage[english]{babel}
\usepackage{caption}
\usepackage{color}

\usepackage{subcaption}

\input glyphtounicode
\ninept

\newcounter{numauth}
\newcounter{listcnt}
\newcommand\authcnt[1]{\ifdefined#1 \stepcounter{numauth} \fi}

\newcommand\addauth[1]{
\ifdefined#1 
\stepcounter{listcnt}
\ifnum \value{listcnt}<\value{numauth}
\appto\authorslist{, #1}
\else
\appto\authorslist{~and~#1}
\fi
\fi}
\authcnt{\paperauthorB}
\authcnt{\paperauthorC}
\authcnt{\paperauthorD}
\authcnt{\paperauthorE}
\authcnt{\paperauthorF}
\authcnt{\paperauthorG}
\authcnt{\paperauthorH}
\authcnt{\paperauthorI}
\authcnt{\paperauthorJ}
\def\authorslist{\paperauthorA}
\addauth{\paperauthorB}
\addauth{\paperauthorC}
\addauth{\paperauthorD}
\addauth{\paperauthorE}
\addauth{\paperauthorF}
\addauth{\paperauthorG}
\addauth{\paperauthorH}
\addauth{\paperauthorI}
\addauth{\paperauthorJ}

\usepackage{times}

\newif\ifpdf
\ifx\pdfoutput\relax
\else
   \ifcase\pdfoutput
      \pdffalse
   \else
      \pdftrue
\fi

\ifpdf 
  \usepackage[pdftex,
    pdftitle={\papertitle},
    pdfauthor={\authorslist},
    pdfsubject={Proceedings of the 25th International Conference on Digital Audio Effects (DAFx20in22)},
    colorlinks=false, 
    bookmarksnumbered, 
    pdfstartview=XYZ, hidelinks 
  ]{hyperref}
  \usepackage[pdftex]{graphicx}
\else 
  \usepackage[dvips]{epsfig,graphicx}
  \usepackage[dvips,
    pdftitle={\papertitle},
    pdfauthor={\authorslist},
    pdfsubject={Proceedings of the 25th International Conference on Digital Audio Effects (DAFx20in22)},
    colorlinks=false, 
    bookmarksnumbered, 
    pdfstartview=XYZ 
  ]{hyperref}
\fi
\usepackage[hypcap=true]{caption}

\title{\papertitle}

\affiliation{
\paperauthorA\ and \paperauthorB\,\thanks{\vspace{-3mm}}}
{IRCAM - CNRS UMR 9912\\Sorbonne Université, Paris, France\\
{\tt \href{mailto:caillon@ircam.fr}{caillon@ircam.fr} | \href{mailto:esling@ircam.fr}{esling@ircam.fr} }
}
\begin{document}
\ifpdf 
  \DeclareGraphicsExtensions{.png,.jpg,.pdf}
\else  
  \DeclareGraphicsExtensions{.eps}
\fi


\maketitle

\begin{abstract}
Deep learning models are mostly used in an offline inference fashion. However, this strongly limits the use of these models inside audio generation setups, as most creative workflows are based on real-time digital signal processing. Although approaches based on recurrent networks can be naturally adapted to this buffer-based computation, the use of convolutions still poses some serious challenges. To tackle this issue, the use of \textit{causal streaming} convolutions have been proposed. However, this requires specific complexified training and can impact the resulting audio quality.

In this paper, we introduce a new method allowing to produce \textit{non-causal streaming} models. This allows to make any convolutional model compatible with real-time buffer-based processing. As our method is based on a post-training reconfiguration of the model, we show that it is able to transform models trained without causal constraints into a streaming model. We show how our method can be adapted to fit complex architectures with parallel branches. To evaluate our method, we apply it on the recent RAVE model, which provides high-quality real-time audio synthesis. We test our approach on multiple music and speech datasets and show that it is faster than overlap-add methods, while having no impact on the generation quality. Finally, we introduce two open-source implementation of our work as Max/MSP and PureData externals, and as a VST audio plugin. This allows to endow traditional digital audio workstations with real-time neural audio synthesis on any laptop CPU.
    
   
\end{abstract}

\section{Introduction}
\label{sec:intro}

Neural audio signal processing has set a new state of art in many fields, such as audio source separation \cite{Stoller2018Wave-U-Net:Separation}, text-to-speech \cite{Ping2018ClariNet:Text-to-Speech}, timbre transfer \cite{Engel2019DDSP:Processing} and unconditional generation \cite{Dhariwal2020Jukebox:Music}. 
Recent works on neural audio synthesis such as DDSP \cite{Engel2019DDSP:Processing}, melGAN \cite{Kumar2019MelGAN:Synthesis} or RAVE \cite{Caillon2021RAVE:Synthesis} have allowed to perform deep audio synthesis faster than real-time. Those methods pave the way towards the integration of neural synthesis and processing inside real-time audio applications. 

Amongst these, models based on recurrent layers (DDSP \cite{Engel2019DDSP:Processing} or RNNoise \cite{Valin2017RNNoise}) are built to process time series sequentially. Therefore, they are naturally fit to process live audio streams by caching their recurrent state in-between DSP calls. However, this is not the case for models based on convolutional networks \cite{Lecun1995ConvolutionalTime-series} since their reliance on \textit{padding} causes audible phase discontinuities between consecutive audio buffers (e.g clicks), which prevents their use for real-time audio applications. A simple solution to address this problem would be to rely on the \textit{overlap-add} method, where we process large overlapping audio buffers and cross-fade them to smooth out phase discontinuities. While this method is straightforward compatible with any generative model, processing overlapping buffers leads to redundant computations and degraded quality during transition phases.
In addition, this method requires caching buffers that are large enough to fill the receptive field of the model in order to avoid edge effects. This results in a high latency between the input and output of the model during inference. A more specific solution have been proposed through the idea of \textit{streaming} models \cite{Rybakov2020StreamingDevices,Zeghidour2021SoundStream:Codec} that use \textit{causal} convolutional layers. These layers replace padding during inference with a cached internal or external state. Although this mechanism allows the use of convolutional models on live audio streams, it usually degrades the model accuracy due to the aforementioned causal constraint.

In this article, we propose a method to make non-causal convolutional neural networks streamable without impacting the audio quality nor introducing computational redundancies.
We achieve this by making the model causal \textit{after training}, leveraging additional internal delays in order to preserve the original computational graph of the model. Hence, our method can be applied over models that were already trained in a non-causal way. As an application case, we use our method to make the recent RAVE model \cite{Caillon2021RAVE:Synthesis} streamable in real-time. However, our approach can be applied straightforwardly to any convolution-based model. We compare our method with several \textit{overlap-add} alternatives using both quantitative and qualitative metrics. We demonstrate that our method outperforms all other baselines in inference speed, while behaving exactly like the original model in terms of audio quality. Finally, we develop several applications leveraging the streaming RAVE model in order to provide regular digital audio workstations with real-time neural audio processing abilities. All of our experiments, methods and source code are packaged as an open-source Python library available online\footnote{\url{https://acids-ircam.github.io/cached_conv}}.


\section{State of art}
\label{sec:soa}




\subsection{Convolutional Neural Networks}

\begin{figure*}[ht]
    \centering
    \includegraphics[width=.75\linewidth]{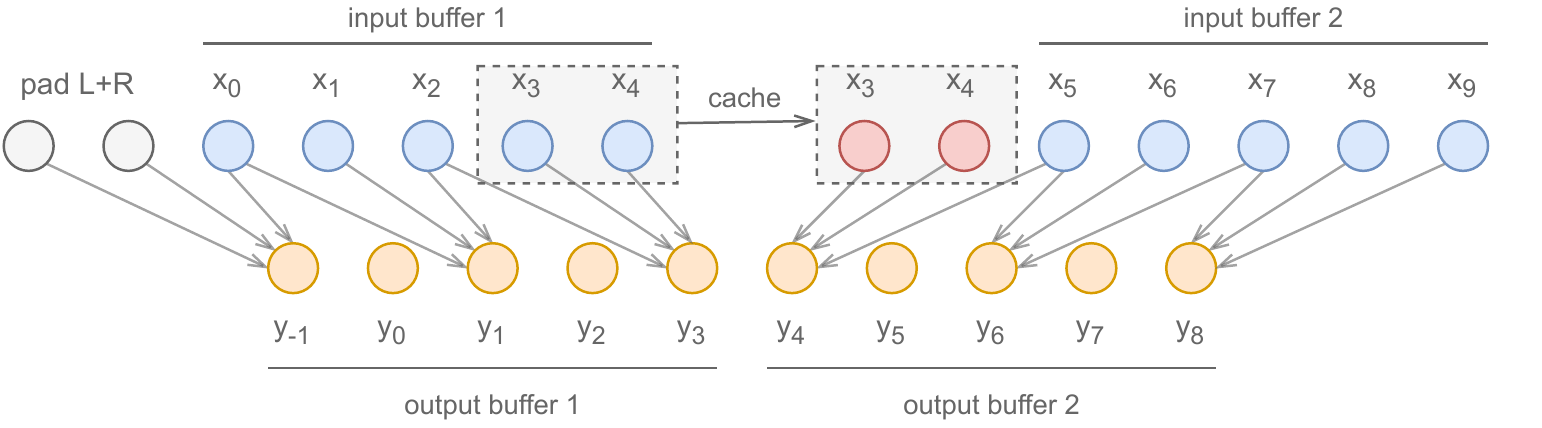}
    \caption{Convolution applied on two split buffers using cached padding. The last $\mathbf N$ frames from input buffer 1 are cached and concatenated with the input buffer 2 (with $\mathbf N$ being the original amount of zero padding) in order to prevent discontinuities between buffers.}
    \label{fig:cached_convolution}
\end{figure*}

We consider a 1-dimensional convolutional layer with a kernel \linebreak$\omega \in \mathbb R^{N\times M\times K}$ applied on an input tensor $x \in \mathbb R^{M\times T}$. The resulting tensor $y$ is defined by
\begin{align}
    \mathbf y^n[i] = \sum_{m=0}^{M-1}\sum_{k=0}^{K-1}\mathbf \omega^{n,m}[k] \mathbf x^m[i + k]
    \label{eq:correlation}
\end{align}
where $ \mathbf y \in \mathbb R^{N \times T-K+1}$.
Due to the size of the kernel $\omega$, the temporal size of $y$ is smaller than the input $x$.
When stacking convolutional layers, this can lead to a significant dimensionality reduction that may be unwanted.
To tackle this issue, convolutional layers are often used in combination with zero-\textit{padding}.
Padding is used to artificially augment the dimensionality of a tensor in order to prevent the loss of dimensionality induced by a convolution with a kernel larger than 1.
As an example, in Equation~(\ref{eq:correlation}), padding the input tensor $x$ with $K-1$ zeros prior to the convolution results in an output tensor $y$ whose temporal dimensionality is the same as the original input.
We call \textit{left-padding} (resp. \textit{right-padding}) the padding of the left-hand side (resp. right-hand side) of the tensor.

Using padding is useful to maintain a tensor dimensionality across layers. However, there are situations where an increase or decrease in temporal dimensionality is required.
Convolutional layers with a stride $s > 1$ allow to decrease a tensor dimensionality by a factor $s$ using the same padding strategy as regular convolutional layers.
On the other hand, transposed convolutional layers can be used to increase a tensor temporal dimensionality.


\subsection{Causal streaming models}
\label{sec:cached_padding}

Processing audio buffers one after the other using a convolutional neural network is not trivial. Indeed, the use of padding in each layer of the model creates discontinuities in the data when processing two consecutive buffers sequentially.
In the context of neural audio synthesis, and more specifically raw waveform modelling, this causes audible phase discontinuities that are not acceptable for real-time audio applications.

To address this problem, Rybakov et al. \cite{Rybakov2020StreamingDevices} proposed to rely on \textit{causal} Convolutional Neural Networks (CNN), which are defined through a \textit{cached padding} mechanism.
%
%
Cached padding is implemented by retaining the end of one tensor and using it to left-pad the following one, as shown in Figure~\ref{fig:cached_convolution}. This allows to maintain continuity between the computation of two consecutive audio buffers. It is meant to be used as a replacement for left-padding during inference, retaining the original padding increase in dimensionality without creating discontinuities in-between buffers. Although this method provides a solution for the use of CNN in real-time audio generation, it is constrained by the necessity to implement \textit{causal convolutions}, which are not widespread. This implies that existing pre-trained models might not be compatible with this method, as most of the existing CNN in the literature do not satisfy this assumption.
Finally, it has been shown that a causal constraint makes the learning process more complex \cite{Rybakov2020StreamingDevices}, which could impact the final audio quality. 

\subsection{RAVE}

The RAVE model \cite{Caillon2021RAVE:Synthesis} is a variational auto encoder \cite{Kingma2014Auto-encodingBayes} applied directly to the raw audio waveform.
It is trained using two separate stages, respectively named \textit{representation learning} and \textit{adversarial fine tuning}.
The representation learning stage uses a spectral distance between the input and output of the model as its main training objective.
The encoder is regularised with a standard Kullback Leibler divergence between the posterior distribution and an isotropic normal distribution.
In order to keep the learned representation as compact as possible, the encoder is only trained during the first stage.
During the second stage, the model is trained using elements from generative adversarial networks \cite{Goodfellow2014GenerativeNetworks} to improve its synthesized audio quality.
A post-training analysis of the latent space is performed as a way to reduce the number of useful latent dimensions.
This allows an easier exploration and manipulation of the latent space.
Overall, RAVE can be used to perform timbre transfer, latent manipulation and unconditional generation with unprecedented quality while synthesizing 20 to 80 times faster than real-time on a laptop CPU. 

RAVE is a feed-forward model, composed of an encoder (a strided convolutional network), and a decoder (a residual transposed convolutional network).
The model also implements the noise synthesizer from the DDSP model \cite{Engel2019DDSP:Processing} to increase its synthesis quality when processing noisy signals.
It leverages zero-padding to maintain the temporal dimensionality of the tensors across convolutional layers.
Therefore, this model in its current state cannot be used to perform streaming inference, and is solely usable on pre-recorded audio files.
Nevertheless, its feed-forward architecture and adversarial fine-tuning makes it a perfect candidate for the streaming task as it is both fast and high quality.

\section{Non-causal streaming models}
\label{sec:post-training-causal}

The streaming models obtained following the method described in Section~\ref{sec:cached_padding} can readily process live audio streams.
However, this requires models that use only causal convolutions, which is not the case for most models proposed in the literature.
Indeed, training a model causally can lead to a loss of accuracy or audio quality \cite{Rybakov2020StreamingDevices}.

Here, we introduce our method that allows to make \textit{non-causal} models streamable.
Our proposal is constructed around the idea of performing a post-training causal reconfiguration of the model. This allows to consider convolutional networks trained using any type of padding (potentially non-causal) and turn them into streamable models. One idea to do so would be to extend the cached padding mechanism to right-padding. However, this is not possible by nature, as we are processing live audio streams where the next buffer is not known yet.

Therefore, we propose to reconfigure the model as causal \textit{after training}. This can be achieved by transforming right-padding into an additional left-padding.
While this reconfiguration allows the use of a cached padding mechanism, making the model causal after training alters its computational graph. Hence, this might produce unpredictable results if the model includes strided convolutions or has a computational graph with parallel branches (e.g residual connections \cite{He2015DeepRecognition}).
In those cases, we propose the introduction of \textit{additional delays} to restore the original behavior of the model. In the following, we detail how we address each of these architectures, in order for our method to be applicable universally on any type of network.




\subsection{Aligning strided convolutions}

\begin{figure}[h]
    \centering
    \includegraphics[width=.6\linewidth]{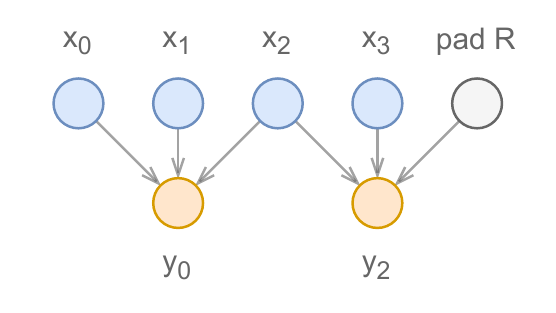}
    \caption{A simplified view of a strided convolution using zero-padding during training.}
    \label{fig:stride_training}
\end{figure}

Strided convolutions are often used as a way to reduce the temporal or spatial dimensionality of an input tensor. This is done by skipping some steps in the application of the convoluted kernel, as depicted in Figure~\ref{fig:stride_training}.

\begin{figure}[h]
    \centering
    \includegraphics[width=.75\linewidth]{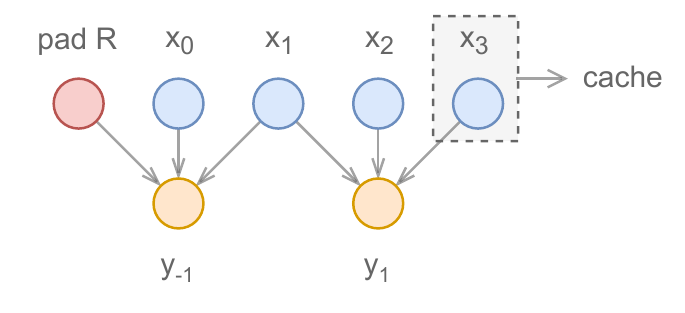}
    \caption{A strided convolution with post-training causal re-configuration. Due to the input lag, the output of the layer is not the same as during training (see Figure~\ref{fig:stride_training} for the regular output).}
    \label{fig:stride_inference}
\end{figure}

Transforming right-padding to left-padding shifts the input tensor to the right (i.e adds a lag to the input tensor). This has no consequence for convolutions with stride 1 or transposed convolutions as it only delays the output tensor.
However, this lag may have an impact on convolutions with a stride greater than one, where a lag of $n$ samples on the input tensor results in a fractional lag of $n/s$ in the output tensor. We show in Figure~\ref{fig:stride_inference} how this fractional lag results in a change of behavior of the layer whenever $n$ is not a multiple of $s$.

\begin{figure}[h]
    \centering
    \includegraphics[width=.85\linewidth]{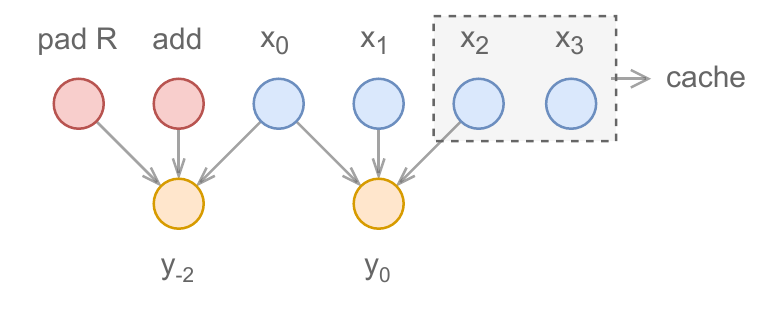}
    \caption{An additional delay (\textit{add}) is applied to the input tensor in order to recover the original behavior of the layer.}
    \label{fig:stride_inference_aligned}
\end{figure}

Therefore, we introduce an additional delay to the input in order to make its overall lag a multiple of the stride of the convolutional layer, as shown in Figure~\ref{fig:stride_inference_aligned}.
In the case of a complex convolutional network, it is necessary to keep track of the overall cumulated lag for an input tensor after each convolutional layer.
Considering that a convolutional layer with stride $S$ and right-pad $R$ processes an input tensor with cumulated delay $D_c$, we need to set the additional delay $D_a$ to
\begin{equation}
    D_a =  S - (R + D_c \mod S) \mod S
    \label{eq:strided_delay}
\end{equation}
This ensures that the overall delay is a multiple of the layer stride.


\subsection{Aligning parallel branches}

\begin{figure}[hbpt]
    \centering
    \includegraphics[width=.8\linewidth]{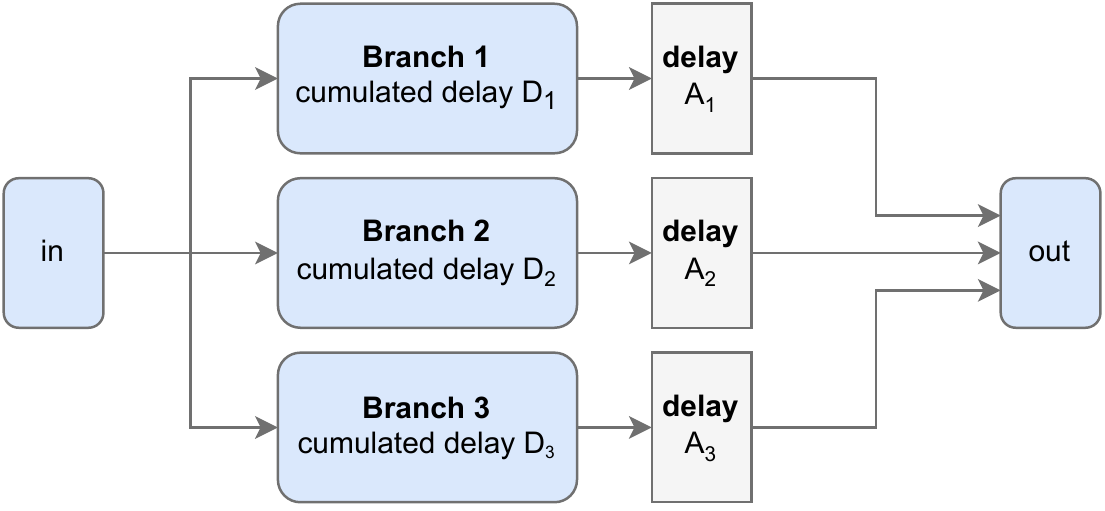}
    \caption{Aligning parallel branches using additional delays.}
    \label{fig:branch_delay}
\end{figure}

When introducing delays inside a computational graph, special care must be given to the alignment of parallel branches. A well-known example of parallel architectures is that of residual layers \cite{He2015DeepRecognition}.
Indeed, residual layers sum the input of a function to its output, in order to make the overall operation act as a perturbation of the identity function.
Hence, it is crucial to delay the residual branch in order to compensate for the delay induced in the main branch by our method enforcing post-training causality. More generally, models implementing parallel branches must introduce delays to re-synchronise the different branches, as shown in Figure~\ref{fig:branch_delay}. In this case, we set the additional delays $A_i$ to
\begin{align}
    A_i = \max_j D_j - D_i,
\end{align}
where $D_i$ is the cumulated delay induced in the $i^\text{th}$ branch.

\subsection{Overlap-add baseline}


For comparison purposes, we use a simple yet effective baseline method to process live audio streams with non-causal convolutional neural networks. We implement the \textit{overlap-add} method by first collecting an audio buffer large enough to account for the receptive field of the model. Then, we apply the unmodified convolutional neural network on this buffer and window the output signal using the Hann window
$$
\mathbf w[n] = \sin \left ( \frac{\pi n}{N} \right )^2,
$$
where $N$ is the buffer size.

\begin{figure}
    \centering
    \begin{subfigure}[b]{.6\linewidth}
        \includegraphics[width=\linewidth]{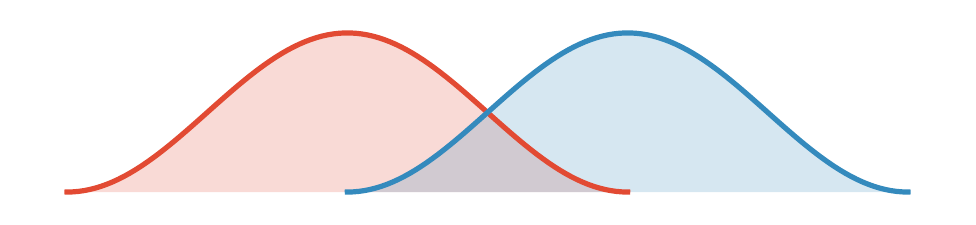}
        \caption{50\% overlap}
    \end{subfigure}
    \begin{subfigure}[b]{.6\linewidth}
        \includegraphics[width=\linewidth]{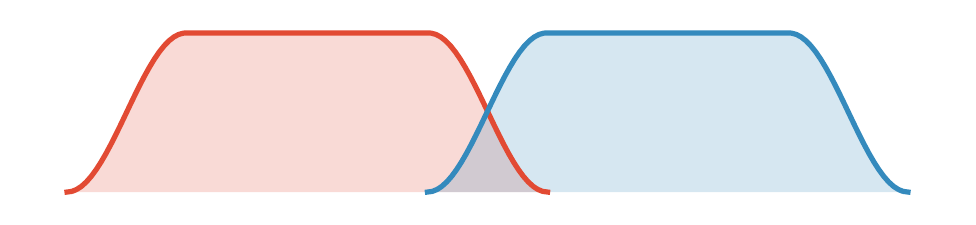}
        \caption{25\% overlap}
    \end{subfigure}
    \begin{subfigure}[b]{.6\linewidth}
        \includegraphics[width=\linewidth]{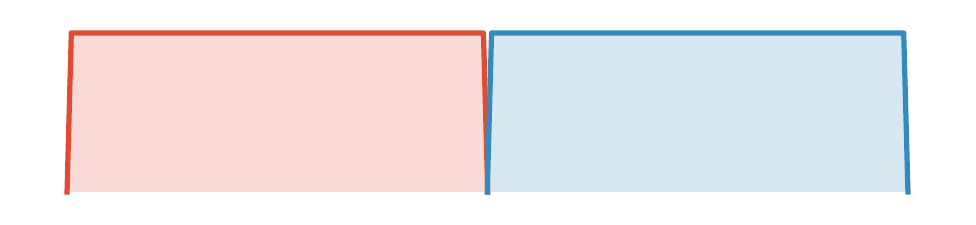}
        \caption{0\% overlap}
    \end{subfigure}
    \caption{Windows used by the three overlap-add baseline variants implemented.}
    \label{fig:overlapping_windows}
\end{figure}

Finally, we add the resulting tensor to the previous output with a temporal offset of $N / 2$. This implements the overlap-add method with a 50$\%$ overlapping factor.
We compare this method to another having a 25\% overlapping ratio, implemented by scaling $w$ accordingly, as depicted in Figure~\ref{fig:overlapping_windows}.
This reduces the computational redundancy of the method and consequently makes it process audio faster. However, using a smaller overlapping window results in harsher transitions between buffers. Hence, we also consider the extreme case of a $0\%$ overlapping factor, where the model is applied on non-overlapping buffers. This last configuration can be seen as an ablation of our method where cached padding and causal constraints are removed.

\section{Evaluation}
\label{sec:exp}

\subsection{Performances}
\label{sec:performances}
In this section, we evaluate the performances of our proposed non-causal streaming method. To do so, we compare it to different variants of the overlap-add method in the context of a model trained without a causal constraint.

In order to evaluate the inference speed, we rely on the Real-Time Factor (RTF) defined as the ratio between processing time and audio duration when processing an audio signal.
A RTF below 1 indicates that the algorithm processes data \textit{faster} than real-time. We also evaluate the amount of memory required during inference on live audio streams, by analyzing the Random Access Memory (RAM) usage. We estimate both memory usage and RTF of the reconstruction process using the various methods applied to 60s long random (white noise) audio signals with varying buffer sizes. We rely on white noise as here the audio output is not relevant to compute the speed of different methods.
All results are averaged over 10 trials in order to account for measurement errors.





\begin{figure*}[ht]
    \centering
    \begin{subfigure}[b]{.42\linewidth}
        \includegraphics[width=\linewidth]{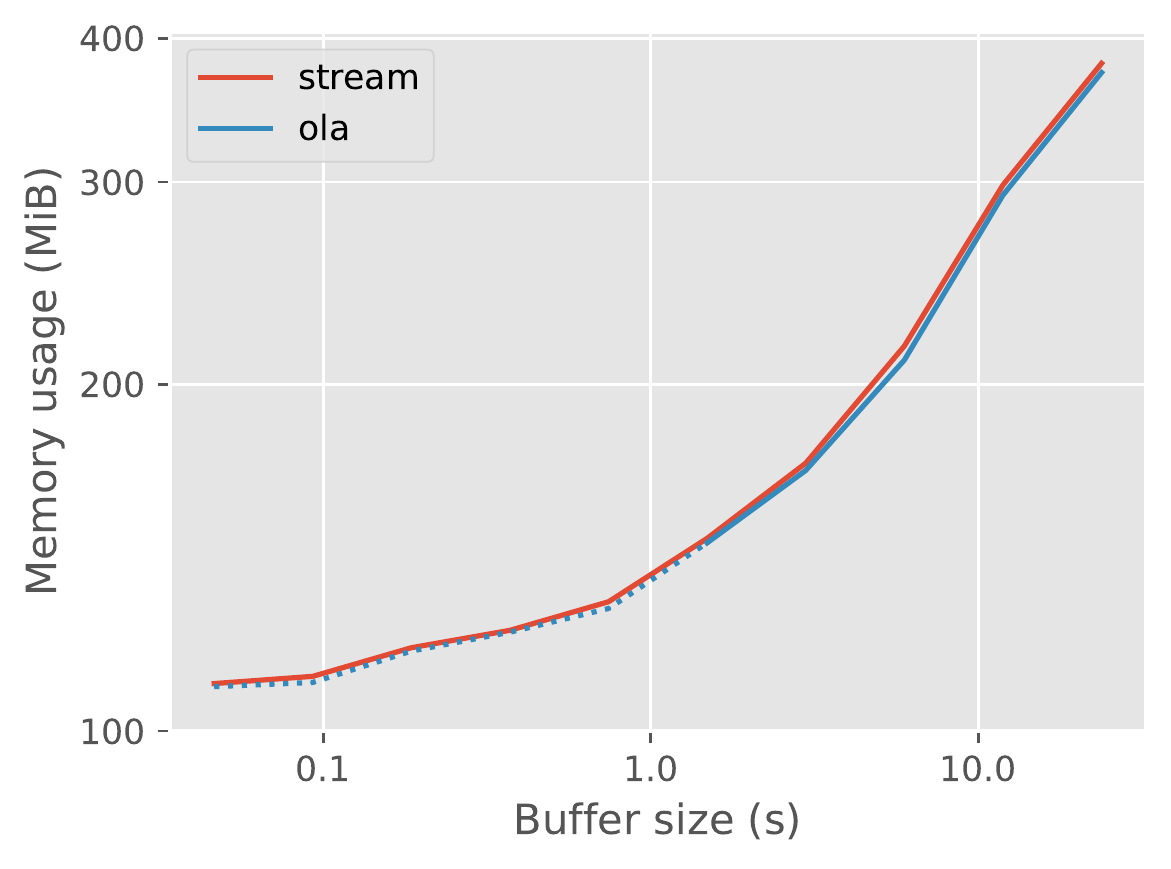}
        \caption{Memory usage}
        \label{fig:memory}
    \end{subfigure}
    \begin{subfigure}[b]{.42\linewidth}
        \includegraphics[width=\linewidth]{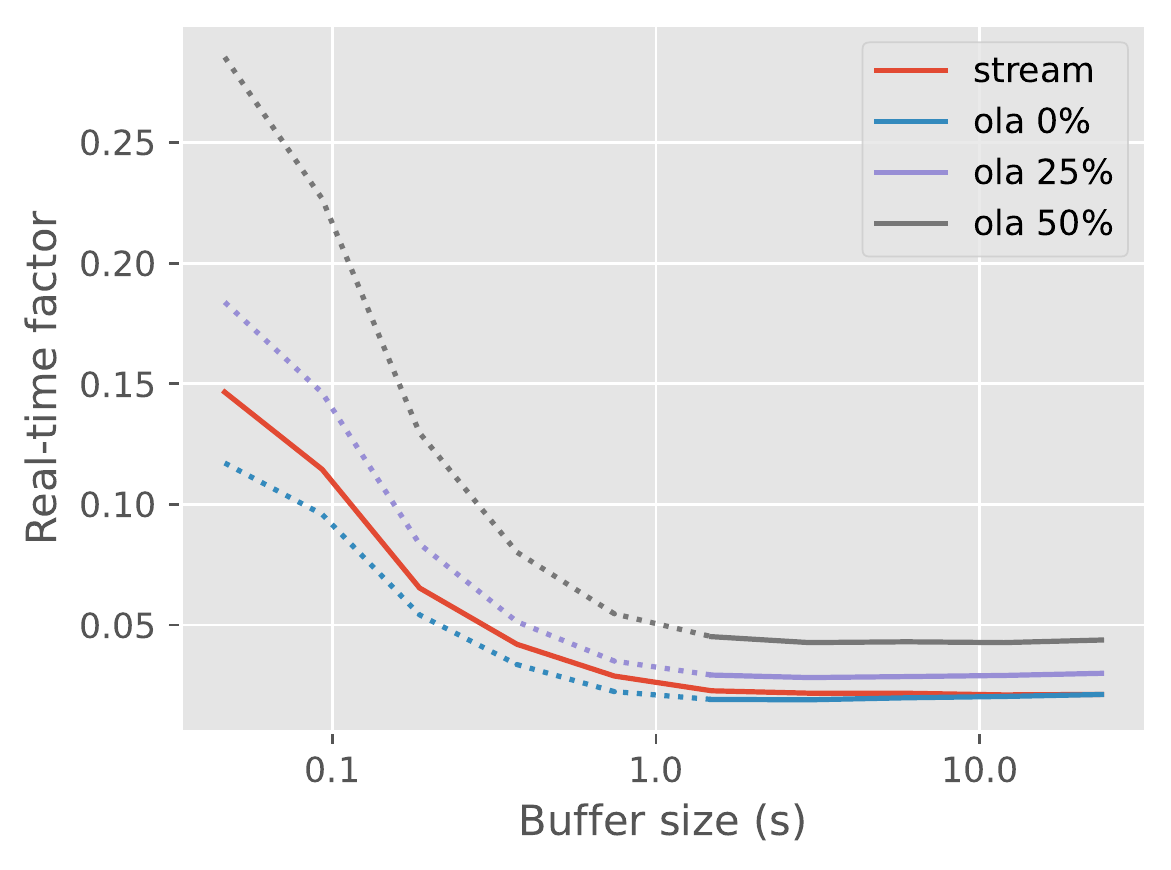}
        \caption{Real-time factor}
        \label{fig:rtf}
    \end{subfigure}
    \caption{Memory usage and real-time factor for the \textit{streaming} and \textit{overlap-add} methods on a regular RAVE model with varying buffer size. Memory usage is identical for all overlap-add methods. Dotted lines indicate that the model is applied on buffers smaller than its receptive field.}
    \label{fig:performances_regular}
\end{figure*}

We show in Figure~\ref{fig:memory} how our proposed \textit{streaming} and different \textit{overlap-add} methods all have a similar memory usage. The only difference comes from a constant 180kiB of additional RAM needed to store the cached padding of the streaming method.

In terms of processing speed, as we can see in Figure~\ref{fig:rtf}, the overlap method with a $0\%$ overlap ratio is the fastest, while also being the less accurate (see Section~\ref{sec:fidelity}). Although increasing the overlap ratio to $25\%$ or $50\%$ can reduce the corresponding artifacts, it also makes the overlap method increasingly slower than the streaming method.
This is due to the computational redundancies involved in this method.

\subsection{Fidelity}
\label{sec:fidelity}

In contrast to our proposed streaming method, the \textit{overlap-add} approach only yields an approximation of the original model.
Hence, we aim to estimate the quality of this approximation by comparing signals coming from the overlap-add method with signals processed offline by a non-causal model.
To do so, we use the two following metrics
\begin{align}
    \mathcal L_{s}(\mathbf x, \mathbf y) &= \| \log (S(\mathbf x)+\epsilon) - \log (S(\mathbf y)+\epsilon) \|_2 \label{eq:spectral_distance}\\
    \mathcal L_{w}(\mathbf x, \mathbf y) &= \| \mathbf x - \mathbf y \|_2 ,\label{eq:waveform_distance}
\end{align}
where $\mathcal L_s$ is a spectral distance computed between amplitude STFT spectrum $S(\mathbf{x})$ and $\mathcal L_w$ is the Euclidean distance between the raw waveforms.
We set $\epsilon=1$ as proposed by Défossez et al. \cite{Defossez2018SING:Generator}.
The spectral distance is useful to assess how perceptually similar two audio signals are, regardless of their phase.
However, the waveform Euclidean distance is highly phase-dependent, and reflects a sample-wise dissimilarity between the raw waveform.
Combined, those two metrics give us insights about how similar signals are both from a perceptual and sample-wise point of view.

We disable the noise synthesizer and set the encoder variance to 0 in order to make the model behave predictably. This is necessary as any randomness involved in the generation process would bias the fidelity measure.

\begin{figure*}[ht]
    \centering
    \begin{subfigure}[b]{.42\linewidth}
        \includegraphics[width=\linewidth]{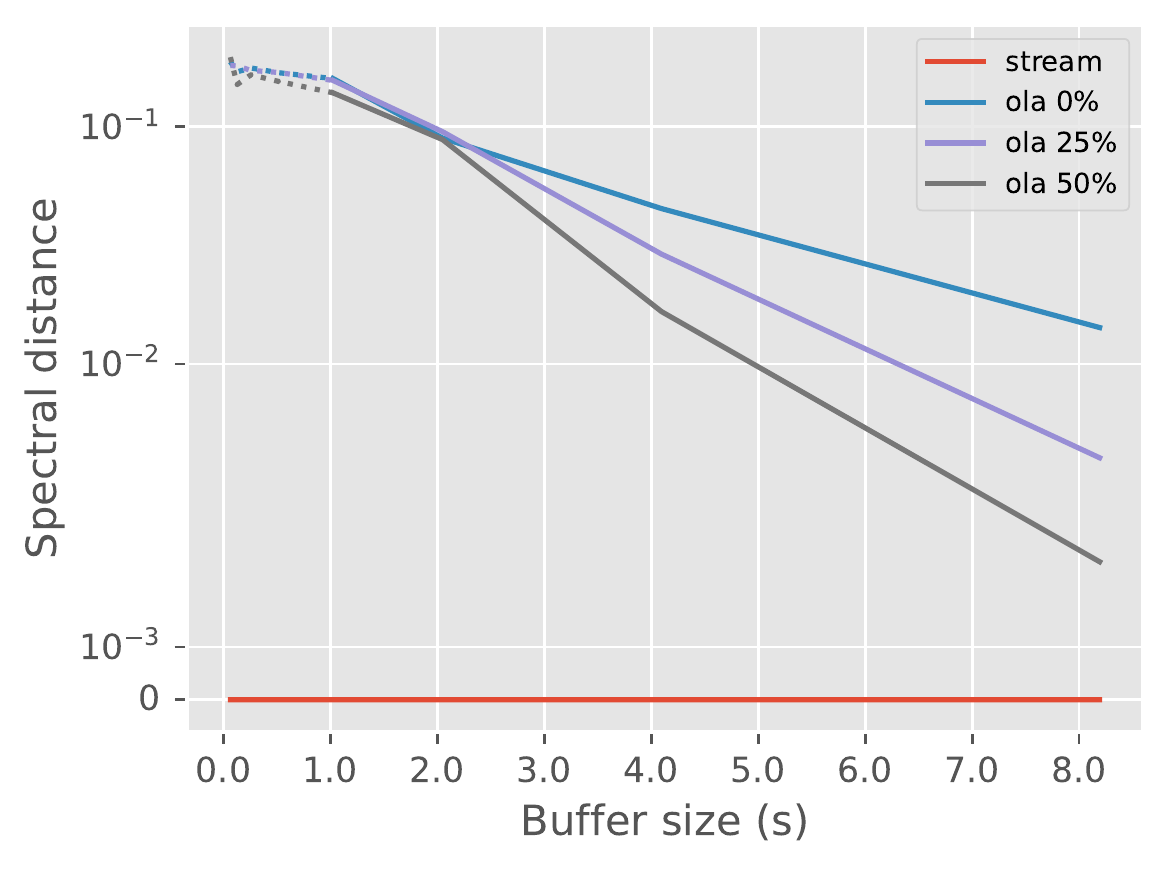}
        \caption{Spectral distance}
        \label{fig:spectral}
    \end{subfigure}
    \begin{subfigure}[b]{.42\linewidth}
        \includegraphics[width=\linewidth]{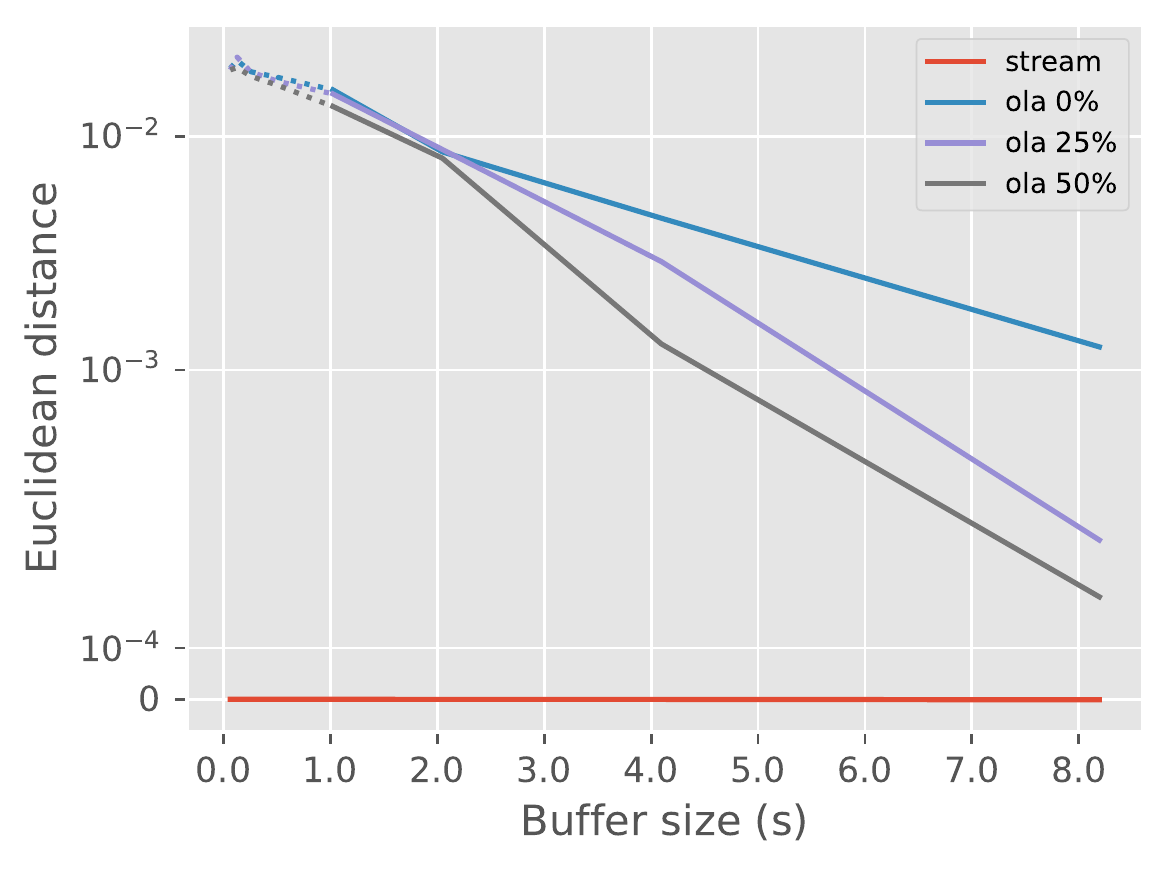}
        \caption{Euclidean distance}
        \label{fig:euclidean}
    \end{subfigure}
    \caption{Spectral and euclidean distances between different \textit{overlap-add} processing methods (ola) and the offline processing method as a function of the buffer size. Dotted lines indicate that the model is applied on buffers smaller than its receptive field.}
    \label{fig:reconstruction}
\end{figure*}

We compare the \textit{overlap-add} methods with several overlapping ratios (0\%, 25\% and 50\%), and also include the \textit{streaming} method to ensure that it is an exact reproduction of the offline method. We compensate the latency present in the synthesized outputs for all methods prior to their evaluation. We test all methods with varying buffer sizes and report the results in Figure~\ref{fig:reconstruction}.

As we can see, all variants of overlap-add methods have a decreasing spectral and Euclidean distances to the offline method as the buffer size increases. However, those distances never become null even for buffer sizes larger than 8s, stressing out the artifacts introduced by such methods. Oppositely, our streaming method is exactly identical to the offline method, regardless of the buffer sizes. This confirms that the cached padding and post-training causal reconfiguration of the model allow its use on live audio streams without altering the quality of the output.


\subsection{Impact of pre-training causal constraint}

As discussed in Section~\ref{sec:cached_padding}, enforcing a causal constraint on the model prior to its training can complexify the modelling task. We evaluate the impact of this constraint on the RAVE model trained with the following internal datasets \\

\noindent \textbf{Darbuka. } It has been shown that modelling percussive sounds using a causal model can be difficult \cite{Huang2018TimbreTron:Transfer}. Therefore, we rely on a dataset composed of various solo darbuka performances sampled at 44.1kHz, with a total duration of approximately 3 hours. \\

\noindent \textbf{Strings. } This dataset contains approximately 30 hours of various strings recordings sampled at 44.1kHz that were scraped from different real-life solo violin performances. Compared to the darbuka, it is composed of harmonic signals with smoother attacks. \\

\noindent \textbf{Speech. } The speech dataset is composed of approximately 8 hours of recordings sampled at 44.1kHz. All recordings are produced by a single speaker in a consistent acoustic environment.\\

All datasets are split into 90\%-10\% validation and train sets. We use all the augmentation strategies proposed in the original article \cite{Caillon2021RAVE:Synthesis}. We train two variants of the RAVE model for each dataset (pre-training and post-training causal re-configuration).
All models are trained for 2M iterations. We use the spectral distance defined in Section \ref{sec:fidelity} to measure the reconstruction error of audio samples from the validation set as input for a pretrained RAVE model. We report the resulting spectral distances in Table~\ref{tab:causal_impact}.

\begin{table}[ht]
  \caption{\itshape Reconstruction errors for pre-training and post-training causal reconfiguration across different datasets.}
	\centering
	\begin{tabular}{c|c|c}
		\hline  ~& pre-training & post-training \\\hline
		Darbuka & $0.228\pm 0.028$ & $\mathbf{0.178\pm 0.038}$  \\
		Strings & $0.055\pm 0.012$ & $\mathbf{0.054\pm0.011}$ \\
		Speech & $0.155\pm0.005$ & $\mathbf{0.138\pm0.005}$ \\\hline
	\end{tabular}
	\label{tab:causal_impact}
\end{table}

Using the pre-training causal configuration results in a small but consistent loss of accuracy as compared to the regular training of models across all datasets.
However, the cumulated lag applied to the input tensor due to the post-training reconfiguration is responsible for a processing latency when using the model on an audio stream.
In the case of the RAVE model, this latency adds up to $653$ms compared to only $52$ms when using RAVE trained with a causal constraint.

\section{Application}

Alongside this article, we also introduce several applications leveraging the streaming RAVE model obtained using our method. This provides real-time neural audio synthesis inside different types of digital audio workstations. The source code and pre-built binaries for all applications are available online\footnote{\url{https://acids-ircam.github.io/cached_conv}}.

\subsection{Max/MSP and PureData externals}

\begin{figure}[ht]
    \centering
    \includegraphics[width=.7\linewidth]{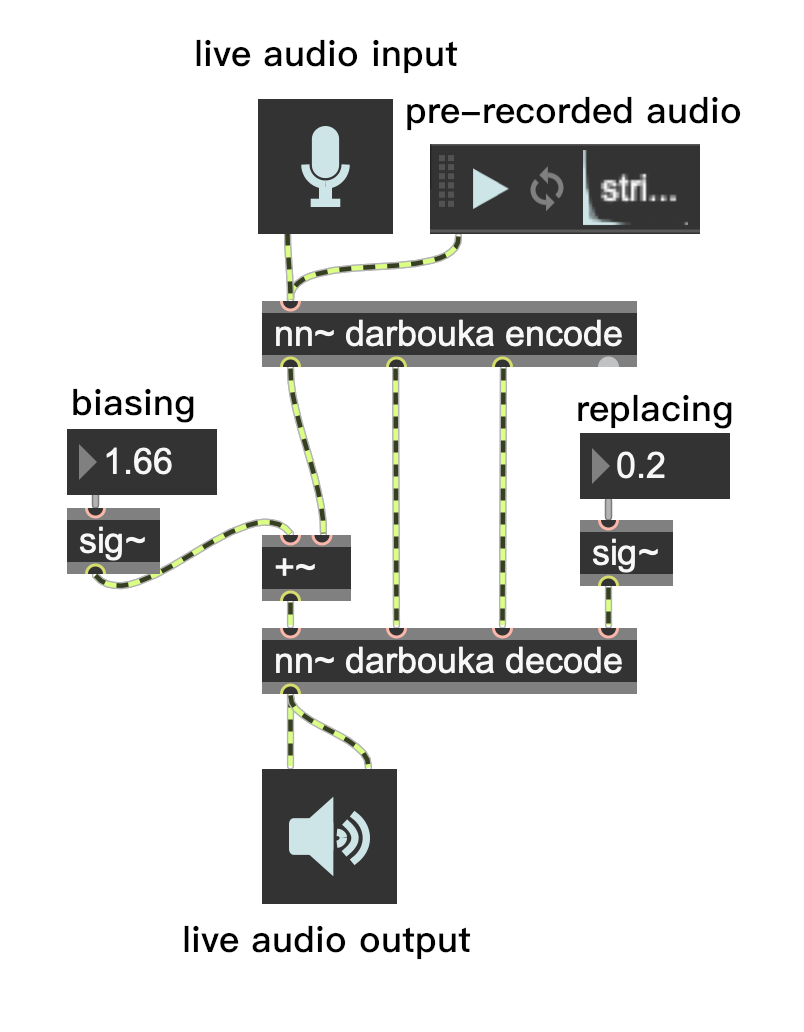}
    \caption{Screenshot of the \textit{nn$\sim$} external interfacing a RAVE model trained on a darbuka dataset. Using either a live audio stream or a pre-recorded audio file as an input, the nn$\sim$ external allows to modify the latent representation yielded by the encoder in real-time. In this example, the first  (resp. last) dimension of the latent space is biased (resp. replaced) by a user defined scalar. }
    \label{fig:nn_tilde}
\end{figure}

We introduce the \textit{nn$\sim$} external for Max/MSP and PureData. This external allows the use of deep learning streaming models to process audio signals inside both applications. It leverages pre-trained models exported as \textit{torchscript} files.
By default, the \textit{nn$\sim$} external uses the \textit{forward} method of the model. However, it is possible to specify another method by passing an additional argument to the external during its initialization. The number of inlets and outlets of the external depends on both the model and the method used. For example, the \textit{forward} method of the RAVE model uses one inlet and one outlet, as both the input and output of this method are monophonic audio signals. However, choosing the \textit{encode} method will create one inlet and $N$ outlets, as the input of this method is a monophonic audio signal, while its output is a $N$-dimensional latent representation. Tensors with a lower sampling rate than audio signals are up-sampled at the audio rate using nearest neighbour interpolation. This method of interfacing $N$-dimensional tensors as audio signals give the user a lot of flexibility, as each individual dimension can be modified in real-time. To examplify this, we show in Figure~\ref{fig:nn_tilde} an example Max/MSP patch where the first and last dimensions of the latent representation yielded by a RAVE model are respectively biased and replaced by a user defined input. 

This implements the high-level manipulation showcased in the original article \cite{Caillon2021RAVE:Synthesis}, but also extended by allowing real-time interaction with the generative process. Overall, the \textit{nn$\sim$} external can be used to combine deep learning streaming models with the large library of objects already available in both MaxMSP and PureData.

\subsection{VST audio plugin}

\begin{figure}[ht]
    \centering
    \includegraphics[width=.85\linewidth]{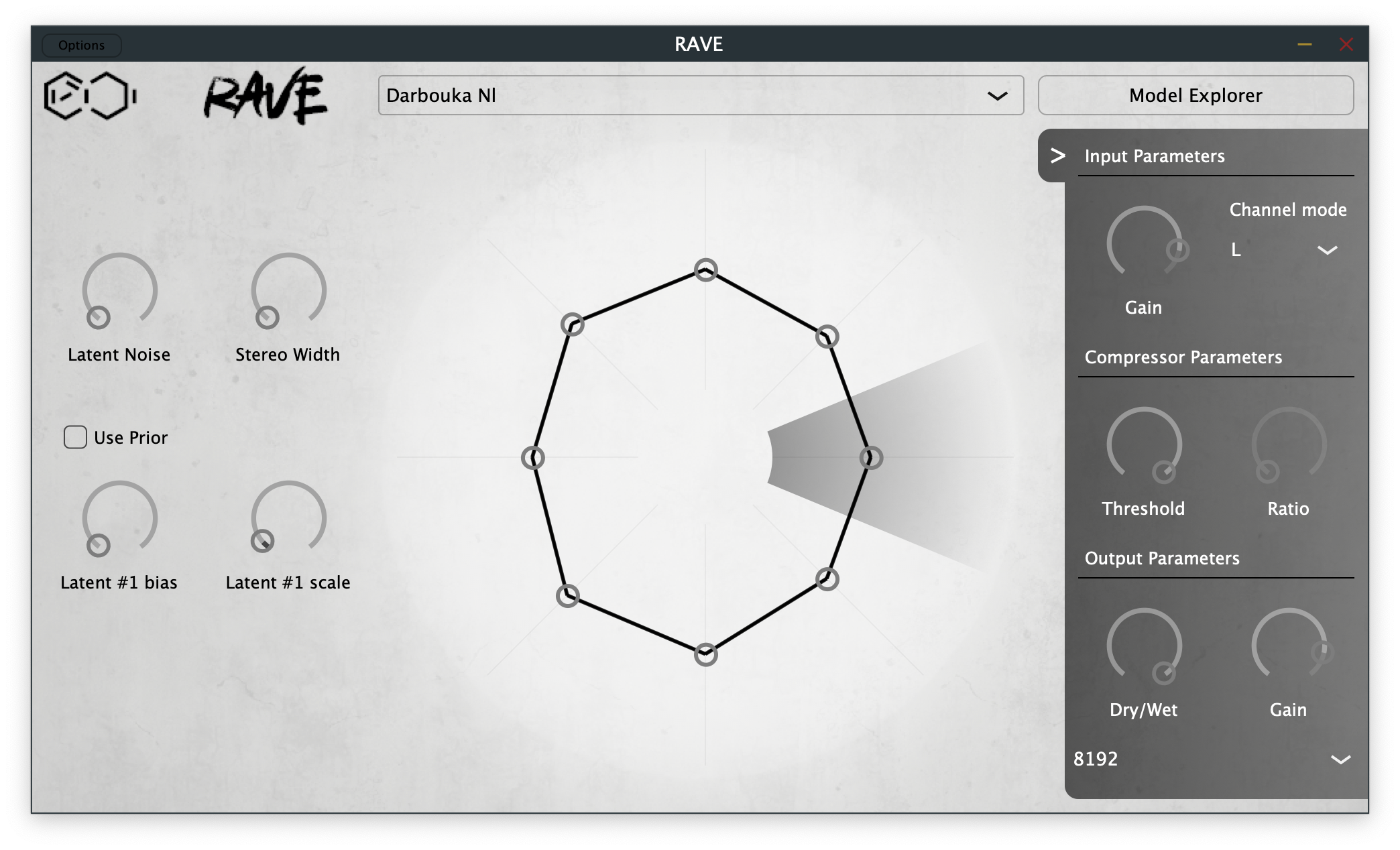}
    \caption{Screenshot of the RAVE VST interfacing a RAVE model trained on a darbuka dataset.}
    \label{fig:vst}
\end{figure}

As an alternative to the \textit{nn$\sim$} external, we propose a VST audio plugin interfacing the RAVE model in order to expand its use to regular digital audio workstations supporting the VST3 plugin format. Our plugin is based on the JUCE framework for both the graphical interface and the audio engine. We depict a screenshot of the plugin in Figure~\ref{fig:vst}.

We generate a latent path, either by using the RAVE encoder on an audio input, or by sampling from the prior of the model. This latent path is then displayed as a circular graph (see Figure~\ref{fig:vst}), where each point corresponds to a single latent dimension. As the latent distribution produced by RAVE is close to a normal distribution, we define the distance $d_i$ of each point from the center of the graph using the following cumulative distribution function

\begin{align}
    d_i = \frac{1}{2}  \bigg(1 + \text{erf}\Big(\frac{z_i}{\sqrt{2}} \Big) \bigg),
    \label{eq:cdf}
\end{align} 

where $\text{erf}$ is the Gauss error function and $z_i$ is the value of the i$^{th}$ dimension of the latent representation. Applying Equation~(\ref{eq:cdf}) to a random variable $x$ sampled from a normal distribution $\mathcal N(0;1)$ results in a uniformly distributed value between 0 and 1.
We give the user the possibility to apply a scale and bias as well as a random noise to each individual dimension. The resulting latent representation is then duplicated and fed to the decoder in order to produce a fake stereo image whose width can be adjusted by the user. We also provide several pre-trained models available in a \textit{model explorer}, where other models will be added over time.

\section{Conclusion and Future Perspectives}

In this paper, we introduced a novel method allowing to transform any convolutional network for audio generation into a streamable model compatible with real-time buffer-based digital signal processing. We showed that our method can be applied on already-trained model by introducing a post-training causal reconfiguration. By carefully handling delays, we showed that this method easily extends to complex architectures with parallel branches. By comparing our method on several speech and music datasets, we showed that it provides faster computation and has no impact on the resulting audio quality. Finally, we released several implementations using our method to provide realtime CNN processing inside digital audio workstations. We hope that this work will pave the way towards the broader integration of the extensive possibilities offered by neural audio synthesis inside creative workflows.

\section{Acknowledgments}
The authors would like to thank Maxime Mantovani for his help on debugging the MaxMSP external and Jean-Baptiste Dupuy and Axel Chemla--Romeu-Santos for their work on the VST audio plugin. This work is currently supported by the ACTOR Partnership funded by the Canadian SSHRC (SSHRC:895-2018-1023) and by the ACIDITeam - Emergence(s) project funded by Ville de Paris.

\bibliographystyle{unsrt}
\bibliography{references}

\end{document}